# Exploring Artificial Intelligence Tutor Teammate Adaptability to Harness Discovery Curiosity and Promote Learning in the Context of Interactive Molecular Dynamics

Mustafa Demir [a], Jacob Miratsky [a, b], Jonathan Nguyen [a], Chun Kit Chan [a], Punya Mishra [c, d], and Abhishek Singharoy [a, b]

[a] Biodesign Institute, Center for Applied Structural Discovery, Arizona State University, Tempe, Arizona, 85281, USA

[b] School of Molecular Sciences, Arizona State University, Tempe, Arizona, 85281, USA

[c] Learning Engineering Institute, Arizona State University, Tempe, Arizona, 85281, USA

[d] Mary Lou Fulton Teachers College, Arizona State University, Tempe, Arizona, 85281, USA

**Manuscript type:** Research Article

**Corresponding Author:**
Mustafa Demir
Biodesign Institute,
Center for Applied Structural Discovery,
Arizona State University
PO Box: Tempe, AZ 85281
mdemir@asu.edu

# ABSTRACT

This study examines the impact of an Artificial Intelligence tutor teammate (AI) on student curiosity-driven engagement and learning effectiveness during Interactive Molecular Dynamics (IMD) tasks on the Visual Molecular Dynamics platform. It explores the role of the AI's curiosity-triggering and response behaviors in stimulating and sustaining student curiosity, affecting the frequency and complexity of student-initiated questions. The study further assesses how AI interventions shape student engagement, foster discovery curiosity, and enhance team performance within the IMD learning environment. Using a Wizard-of-Oz paradigm, a human experimenter dynamically adjusts the AI tutor teammate's behavior through a large language model. By employing a mixed-methods exploratory design, a total of 11 high school students participated in four IMD tasks that involved molecular visualization and calculations, which increased in complexity over a 60-minute. Team performance was evaluated through real-time observation and recordings, whereas team communication was measured by question complexity, and AI's curiosity-triggering and response behaviors. Cross Recurrence Quantification Analysis (CRQA) metrics reflected structural alignment in coordination and were linked to communication behaviors. High-performing teams exhibited superior task completion, deeper understanding, and increased engagement. Advanced questions were associated with AI curiosity-triggering, indicating heightened engagement and cognitive complexity. CRQA metrics highlighted dynamic synchronization in student-AI interactions, emphasizing structured yet adaptive engagement to promote curiosity. These proof-of-concept findings suggest that the AI's dual role as a teammate and educator indicates its capacity to provide adaptive feedback, sustaining engagement, and epistemic curiosity. Future research will refine AI strategies by integrating multimodal data to enhance curiosity-driven learning.

# INTRODUCTION

**Artificial Intelligence as a Teammate**

Presently, Artificial Intelligence (AI), particularly generative Large Language Models (LLMs, e.g., ChatGPT; *ChatGPT*, n.d.), transforms the education sector by introducing AI as a collaborative "teammate" that goes beyond traditional teaching methods and materials (Demir et al., 2024). Unlike conventional machines, defined as "a device, having a unique purpose, that augments or replaces human or animal effort for the accomplishment of physical tasks" (Rafferty, 2024, *para*. 1), AI transcends this by engaging in complex cognitive tasks, adapting to human interactions, and fostering epistemic curiosity through real-time, adaptive feedback in educational settings.

Furthermore, AI's versatility in adapting to individual learning needs also fosters the potential to support the development of essential metacognitive skills, such as critical thinking and problem-solving, that are needed for handling complex learning tasks. Additionally, AI can support shared mental modeling, which is the organized mental representation of the key elements within the team's environment that are shared across team members (Klimoski & Mohammed, 1994), particularly when thoughtfully integrated into team task domains. The development of these skills is not an autonomous function of AI; rather, it depends on effective human-AI collaboration and teaming. Without such thoughtful integration, there is a notable risk that AI could encourage cognitive outsourcing, thereby undermining the very metacognitive abilities it has the potential to support (Yan et al., 2024). These metacognitive skills are arguably more beneficial and impactful when developed and applied within such collaborative settings, as they underpin effective teamwork, facilitate the leveraging of diverse perspectives, and are crucial for navigating the increased complexity of team-based learning tasks. For instance, when utilized as a component of human-led strategies, AI as a tool can serve teams by monitoring and assisting their interactions, accessing varied information, or even offering prompts that encourage reflection on their collaborative strategies (Willett & Demir, 2023). However, the Shein (2024) study underlines that while AI can accelerate aspects of problem-solving, foundational understanding and the metacognitive processes that build it remain essential, perhaps even more so at the group level (Wu & Or, 2025). In this case, students who utilized AI tools to solve programming tasks, whether individually or as part of a team, performed well initially but demonstrated reduced retention over time. This underscores the critical importance of guidance, human oversight, and structured opportunities for reflection within the team learning process, ensuring that AI tools function as genuine supports within a collaborative human-AI framework rather than inadvertently undermining the development of deep, collective understanding and sustaining long-term learning outcomes for the group.

Even though AI models have advanced this much, the U.S. Department of Education's 2023 policy underscores a "humans-in-the-loop" approach, where teachers are critically important in guiding and contextualizing AI-driven insights, which ensures AI's adaptability aligns with pedagogical teaching task goals while mitigating risks of algorithmic bias, a.k.a. hallucinations (Emsley, 2023), and over-reliance and trust in automated judgments (Tossell et al., 2024). This need for guided, trustworthy AI is particularly acute in K-12 STEM education, where fostering early engagement and providing scalable support in complex scientific subjects can significantly influence student pathways. AI systems enable tailored, real-time communication and coordination, creating individualized learning experiences that dynamically adjust to each learner's level of curiosity, skill, and the speed at which they progress through educational material. Rather than replacing teachers, AI enhances their roles, allowing educators to concentrate on complex instructional strategies and facilitating a more personalized learning experience (Lakhani, 2023). Such an approach not only preserves the intentionality of learning but also emphasizes the depth required for effective education, with educators remaining in a central role in decision-making and instructional guidance (U.S. Department of Education, 2023).

In a human-in-the-loop framework, conceptualizing AI as a team member leads to the concept of Human-AI Teaming (HAT), a sociotechnical system where humans and intelligent machines interact interdependently to adapt to evolving tasks and work collaboratively to achieve shared goals (Demir, Cohen, et al., 2023; Demir & Cooke, 2014). This HAT framework draws on principles from the broader fields of team science and team cognition, which examine how interdependent individuals with heterogeneous task roles coordinate their actions and thinking in collaborative problem-solving (Wiltshire et al., 2017) and learning environments (Grand et al., 2016). A key element of effective teamwork is the development of shared mental models, wherein team members engage in effective team interactions (i.e., communication and coordination) and effective decision-making, have a mutual understanding of the overall team task, and recognize each other's taskwork (Demir, Canan, et al., 2023).

Consequently, for effective teamwork, HAT requires characteristics such as role interdependence, independence in achieving task goals, and enhanced performance in dynamic environments through effective interactions, where AI can participate as a team member. This team interaction dynamics is enhanced when learners develop shared mental models with the AI, allowing for more intuitive information foraging and deeper engagement with complex concepts. In this context, AI serves as a collaborative team member that strengthens foundational knowledge and actively cultivates adaptive skills and epistemic curiosity, both of which are essential for success in an increasingly AI-driven workforce.

**Epistemic Curiosity via AI Tutor Teammate**

As intelligent machines transition from tools to teammates, the role of AI in fostering curiosity becomes increasingly significant because of the complex nature of epistemic curiosity, where AI can shape how students explore, question, and master knowledge. Therefore, considering AI as a collaborative teammate in education not only transforms the nature of learning environments but also directly influences the ways in which students obtain and engage with new information. AI's omnipresence is already rich in epistemic curiosity, which is an emotional-motivational process parallel to learning, to acquire new knowledge that *motivates individuals to learn new ideas, eliminate information gaps, and solve intellectual problems* (Litman, 2008; Litman & Jimerson, 2004; Litman & Spielberger, 2003; Zedelius et al., 2022). This process is particularly powerful in human-AI interactions because it operates at both individual and social levels, where the AI can simultaneously act as a trigger for cognitive conflicts and a scaffold for meaning-making during discovery. This aligns with theories of conceptual change, where actively creating and guiding learners through the resolution of cognitive conflict is a powerful pedagogical strategy for sparking curiosity and fostering deeper understanding.

Even the most advanced learners often seek an instructor's explanation to understand complex foundational ideas and concepts in a discipline (Hoffman et al., 2018), which can be a barrier to asking productive questions, hinder deeper learning, and require epistemic curiosity. Thus, a learner's epistemic curiosity can be activated when there are expectations of discovery opportunities for learning new knowledge (via the discovery or gathering of new information) that would generate pleasurable experiences of situational interest (Mishra, 2022). This general tendency to delight in gaining new information leads to greater general knowledge, enhanced responsiveness to new information, and superior ability to distinguish between real and made-up concepts, all complemented by a penchant for intellectual humility (Litman, 2008; Mishra, 2022; Zedelius et al., 2022). To design effective educational technology, fostering curiosity is a central goal, particularly when aiming to enhance discovery-based learning.

Two prominent theories underpin the understanding of how curiosity is triggered and maintained in learners. The first is the *Information Gap Theory* (Berlyne, 1960), which proposes that curiosity is stimulated when individuals recognize a discrepancy between their current knowledge and the information they desire to acquire. Accordingly, this gap creates an intrinsic motivation to seek and obtain new knowledge, thereby closing the perceived gap and driving engagement and learning. This theory is particularly relevant in educational environments where learners are encouraged to explore topics that challenge their existing understanding, motivating them to actively seek and obtain new information to fill that gap. The presence of AI as a teammate creates a unique dynamic where information gaps can be identified and bridged in real-time, allowing for adaptive scaffolding that maintains optimal challenge levels for sustained curiosity.

Second, the *Theory of Arousal* (Loewenstein, 1994) suggests that curiosity can also be driven by novel and complex stimuli that heighten an individual's arousal. According to this theory, when learners are presented with unfamiliar, challenging, or ambiguous content, their curiosity is piqued, leading them to engage more deeply with the material. Therefore, the arousal caused by encountering something new or unexpected compels individuals to explore further in order to make sense of its complexity. This theory emphasizes the importance of designing educational experiences that are stimulating enough to capture attention and maintain curiosity over time. In AI-supported learning environments, arousal can be precisely controlled through dynamic difficulty adjustment and cognitive load management, creating an optimal flow state, i.e., a mental condition characterized by complete focus on a single task or activity (Csikszentmihalyi, 1990), where curiosity and challenge are effectively balanced. Thus, the intersection of these two theories provides a valuable framework for developing AI-powered educational technologies that identify areas where learners lack knowledge (Information Gap Theory) and present these topics in compelling and motivating ways (Theory of Arousal).

To understand this impact, it is essential to examine the dynamics of curiosity itself, especially in the context of how different forms of curiosity, i.e., discovery and deprivation, play out in learning scenarios. While "discovery curiosity" generates a positive effect when learning something new towards exploratory or mastery-oriented learning (Litman, 2008; Mishra, 2022; Zedelius et al., 2022), it also creates powerful moments of metacognitive awareness where learners actively construct meaning from their interactions. This contrasts with "deprivation curiosity,"

In models such as Litman's (2008) "deprivation" refers to the unsettling cognitive state experienced when an individual recognizes a specific gap in their knowledge; this perceived lack, in turn, triggers deprivation-type curiosity, an urgent drive to reduce uncertainty through targeted exploration. This drive often channels learning towards performance-oriented goals, where the primary aim is to acquire the specific information needed to resolve the immediate cognitive discomfort. However, this intense focus on rapidly resolving a specific informational deficit is also associated with particular cognitive downsides, such as errors, confusion, a lack of intellectual humility, and increased susceptibility to misinformation, including fake news (Zedelius et al., 2022). Theoretically, such outcomes may arise because the urgency inherent in deprivation curiosity can lead to a narrower and less critical search for information, prioritizing the speed of gap-filling over a thorough evaluation of sources. This focus on performance over mastery can result in more superficial processing, making individuals more vulnerable to accepting plausibly sounding but incorrect information.

In a team context, epistemic curiosity and learning can be indicated by who knows what within a team (Wilson et al., 2007); they can be achieved via interaction, i.e., coordination of information exchange among the agents. To this end, in engineering, it is assumed that AI also has some level of curiosity, which alters maximizing the AI's learning about itself and its team task environment by selecting optimum-level actions. In this manner, AI's artificial curiosity is "driven" by intrinsic motivation for maximum learning that engages reinforcement learning, in which intrinsic rewards are proportional to their learning progress (Epstein & Gordon, 2018; Gordon et al., 2015). Compared to an individual level, the HAT information-seeking and collaborative problem-solving (Wiltshire et al., 2017) behaviors that emerged from interactions among the team members might be more effective in obtaining new knowledge to minimize or eliminate uncertainty. In this case, "discovery curiosity" at an individual and team level during problem-solving and their learning outcomes are influenced by team coordination dynamics to obtain new taskwork or teamwork knowledge (Ravari et al., 2021). Effective team learning requires establishing a dialogical space between the team members to balance team coordination, i.e., stable and flexible coordination dynamics at some level, leading team members to understand the problem, evaluate possible solutions, and construct knowledge (Demir et al., 2019).

To better understand this balance in HAT, the current study explores how the AI team members can leverage these theories to foster *discovery curiosity* in learners. Applied Interactive Molecular Dynamics (IMD) tasks conducted within the Visual Molecular Dynamics (VMD) software platform are often characterized by their complexity and novelty (Humphrey et al., 1996), making them well-suited to activating the mechanisms outlined in both the Information Gap and Arousal theories discussed above. The AI tutor teammate can be designed to dynamically present tasks and challenges that reveal learners' knowledge gaps while offering stimulating and novel experiences. By doing so, the AI would not only respond to learner queries but actively guide and regulate curiosity by moderating the difficulty and novelty of tasks, ensuring maintained engagement. This research question opens avenues to investigate how AI can adaptively frame the learning process, identifying moments when learners are on the cusp of discovery curiosity, and using AI interventions to maximize curiosity-driven exploration in the IMD tasks conducted within the VMD software.

Building on this foundation, we introduce a practical framework to operationalize how an AI-tutor teammate can foster curiosity in learners: a "push-pull" communication strategy that dynamically balances challenge and support. This approach is grounded in the Information Gap and Arousal theories, translating them into real-time interaction mechanisms. In this dynamic, "curiosity triggers" function as the push, where the AI proactively introduces intriguing questions, unexpected challenges, or novel facts. These are designed to spark information gaps and elevate arousal levels, prompting exploration. Conversely, "curiosity responses" operate as the pull, offering scaffolded explanations that help learners bridge gaps, solidify understanding, and reduce cognitive load. We propose that the effective orchestration of this push-pull interplay is a central mechanism for sustaining curiosity-driven learning. By aligning AI interventions with learners' cognitive-emotional states, this strategy not only nurtures discovery curiosity but also promotes metacognitive engagement and effective learning within HATs.

**Synchronization in HAT**

The addition of AI as a collaborative teammate to team complex tasks requires a deeper examination of how human-AI collaboration can be optimized, particularly through synchronization, to create more responsive and tailored learning experiences (World Economic Forum, 2024). Synchronization refers to the alignment of rhythms between interacting systems due to weak coupling, resulting in coordinated behavior (Pikovskij et al., 2003). That is, when coupled, self-sustained oscillators differ from harmonic oscillators by locking onto a single frequency instead of producing a mixture of all mode frequencies, resulting in synchronization (Strogatz, 2015).

In HAT, this synchronization can be seen as the dynamic coordination of rhythms, actions, and decision-making between human and AI tutor teammates, achieved through continuously adjusting behaviors based on real-time interactions and feedback to accomplish a shared task goal. This process arises from what is termed "weak coupling" in this context, meaning the interaction between human and AI should be supportive but not dominating, allowing both agents to maintain their individual decision-making abilities while still adjusting to each other in a coordinated manner. It is through this flexible coupling that strong, synchronized behavior can emerge.

This synchronization in HAT is multifaceted, involving alignment across several dimensions that go beyond simple behavioral mimicry, with key components include: (1) *Mutual adjustments:* Both human and AI tutor teammates adjust their actions based on the other's behavior, similar to the concept of phase locking, where rhythms are adjusted to achieve alignment. This co-regulation process creates a dynamic learning space where both partners can operate within their optimal zones of proximal development; (2) *Information flow:* Continuous feedback loops where both human and AI tutor teammates exchange information ensure that the team stays synchronized in terms of task priorities. This creates a form of distributed cognition where the strengths of both human intuition and AI processing can complement each other; (3) *Goal alignment*: Synchronization ensures that the human and AI work towards a shared goal or task, with their activities timed and coordinated in a way that promotes effective teamwork. Therefore, HAT coordination is a subset of synchrony that involves a deeper alignment in the timing and nature of team cognitive processes beyond just coordinating tasks. It is an ongoing process that ensures the team remains adaptive, responsive, and aligned in achieving common goals or tasks. A central premise of this work is that the quality of this dyadic synchronization is a key factor in creating an interactive environment conducive to fostering epistemic curiosity.

Therefore, to address these challenges within the K-12 context, particularly focusing on high school students, this study investigates discovery curiosity and learning effectiveness using qualitative and quantitative approaches to contribute to the understanding of AI-powered educational technologies. We specifically draw on prior research, including modified survey questions from Litman and colleagues (2008), to examine how students' epistemic curiosity changes before and after interacting with an AI-tutor teammate at the K-12 level in applied Interactive Molecular Dynamics (IMD) tasks conducted within the VMD software platform. To further understand the dynamics of student interactions over time, we employ nonlinear dynamical systems (NDS) methods, exploring how knowledge acquisition unfolds and how it relates to learning effectiveness in applied molecular dynamics tasks. To answer our research questions, our analysis is structured as follows: first, we examine the concept of synchronization between the AI-tutor teammate and each student by extracting NDS metrics; then, we discuss the relationship of each of these metrics with curiosity and learning outcomes. Finally, based on the findings, we underline the cognitive design requirements for developing more adaptive AI edtech systems, emphasizing the need to account for individual differences in effective teaching practices.

## CURRENT STUDY

**Visual Molecular Dynamics Software Platform**

As part of a larger project (Leahy et al., 2023), this study examines human-systems integration from the practitioner perspective, focusing on developing a collaborative AI tutor teammate to foster discovery curiosity and enhance learning effectiveness. Accordingly, the AI tutor teammate guides participants through interactive molecular dynamics (IMD) tasks within the VMD software, a platform for visualizing and analyzing molecular structures (Humphrey et al., 1996). In molecular dynamics or MD simulation, every atom of the system (protein, membranes, DNA/RNA, solvent, and ions) is represented with the dynamics arising from the application of Newton's Second Law combined with a complex energy function (Hollingsworth & Dror, 2018). Visual Molecular Dynamics, also known as VMD, enables users to explore complex molecular systems, visualize various molecular representations, and interact in real-time with simulations for conducting IMD tasks (Stone et al., 2001). IMD is useful because it allows researchers to directly manipulate and observe molecules within a simulation in real-time, providing a more intuitive understanding of molecular behavior and enabling exploration of complex systems by "feeling" how molecules interact, which can lead to new insights and hypotheses that might not be apparent through traditional simulation methods alone; particularly valuable for studying protein dynamics, drug-protein interactions, and the behavior of molecules within channels or membranes.

In this study, we trained the AI tutor teammate model with visual and text recordings of molecular movements captured within the VMD software. By providing guiding questions (i.e., pulling of information) and real-time constructive feedback (i.e., pushing of information), the AI supports learners in comprehension and retention of abstract biophysical concepts during the IMD task.

Figure 1 illustrates a comprehensive VMD interface utilized by the student during an IMD session, highlighting the interactive exchange between the student and the AI tutor teammate. The main visualization (Figure 1a) displays a molecular structure with specific residues labeled (THR79 and ASP36), highlighting their spatial relationship. A ChatGPT dialog box (Figure 1b) captures a dynamic information exchange between an AI tutor teammate and a student, where the AI tutor teammate guides the student by encouraging exploration and deeper understanding of the molecular structure displayed. For instance, the AI tutor teammate guides students through tasks, measuring distances between residues, interpreting visual representations, and exploring secondary structures, such as alpha helices and loops. Through curiosity-triggering questions and responses, the AI-tutor supports the student's learning process, helping them engage actively with the molecular visualization tasks within VMD's scientific environment. Supporting the main visualization, other interface panels include a command log (Figure 1c), which displays molecular and atomic details; graphical representation settings (Figure 1d) for adjusting molecular views, and a molecule selection and control panel (Figure 1e), all of which aid in a detailed molecular exploration. The TkConsole (Figure 1f) facilitates command input, allowing customized molecular analysis.

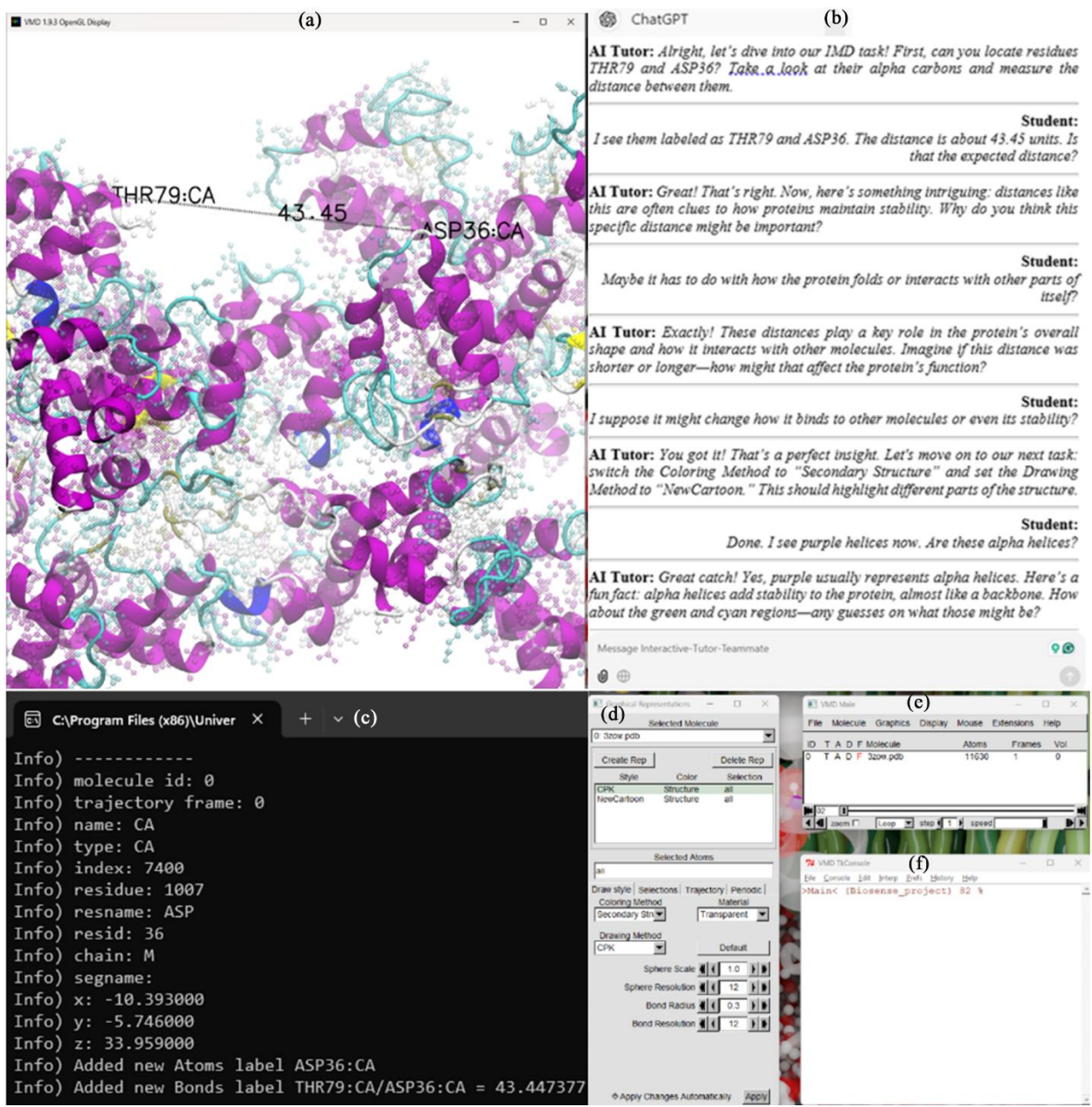

**Figure 1. (a)** VMD interface displaying molecular structure visualization, **(b)** ChatGPT dialog box guiding the IMD tasks with AI tutor teammate and student interaction, **(c)** command log with molecular and atomic details, **(d)** graphical representation settings, **(e)** molecule selection and control panel, and **(f)** TkConsole for command input for the BioSense project.

**Simulated Interactive Molecular Dynamics Task and Experimental Design**

IMD tasks are designed to engage learners actively, allowing them to explore molecular interactions and dynamic processes, which can stimulate curiosity and deepen understanding (O'Connor et al., 2019; Stone et al., 2001). The experiment employed a *Wizard of Oz* (WoZ) approach (Riek, 2012), where the experimenter and the student interacted via text-chat-based communication on Zoom while being introduced as the AI tutor teammate. This approach allows the experimenter to interact with participants while simulating AI behavior, thereby maintaining the appearance that participants are communicating with an autonomous AI system. The WoZ setup enabled the experimenter, acting as the AI, to maintain optimal cognitive disequilibrium and foster an effective interaction with each student. In this case, the experimenter would utilize ChatGPT-4o to provide clarifying explanations, offer alternative strategies, or adjust the task's difficulty to sustain curiosity-driven engagement. Instances where intervention was considered necessary include when a student was unable to proceed with a task due to insufficient scripting support, to clarify ambiguous guidance that was confusing, or to correct a misdirection that was impeding progress, such as when students struggled with protein solvation or locating a specific structure.

In this role, the experimenter utilizes ChatGPT4o to answer participants' questions or to assist in task completion, creating a dynamic feedback loop that enhances the learning experience. This setup ensures a controlled environment where participants can explore complex tasks while the experimenter (acting as the AI) adjusts communication and task difficulty to promote curiosity-driven engagement. The AI-Tutor Teammate communicated and communicated with each student by *curiosity trigger* (pulling of information), intriguing facts, and posing challenging questions to spark curiosity and prompt student inquiries; *curiosity response* (pushing of information) offers deeper insights and suggests further exploration based on students' questions.

In a total 60-minute task duration, participants, with guidance from an AI tutor teammate, were asked to complete an IMD task that involved analyzing the properties of Cytochrome c (Cytc) by using the VMD software. Due to Cytc's well-established role in essential cellular functions and its suitability as a model for exploring molecular properties' characteristics (Dickerson & Timkovich, 1975), such as solvation behavior, surface exposure, and electric charge distribution, provides a focused yet rich framework for real-world analysis. So, for a structured learning task, Cytc can provide meaningful insights into molecular interactions with guidance from an AI-tutor, thereby bridging complex concepts in an accessible, real-world context.

The task was divided into four main components in an ordered manner, as summarized in Table 1. Accordingly, participants engaged in four tasks with Cytc, progressing from basic to more complex computational and theoretical challenges within the context of IMD. To do so, participants selected and justified a unique molecular representation, starting with solvation and visualization. Next, they calculated the number of water molecules within a specified cutoff distance around various residue types (polar, nonpolar, and charged). They then computed the Solvent Accessible Surface Area (SASA) of Cytc and analyzed its connection to the previous water molecule calculation. Finally, participants calculated and plotted Cytc's dipole moment, exploring its functional implications. As the tasks intensified, participants were encouraged to interact dynamically with the AI tutor teammate, asking questions and receiving real-time feedback, enhancing their understanding and problem-solving skills.

**Table 1.** Experimental tasks to assess Cytc modeling skills and AI tutor teammate collaboration, from visualization to advanced coding.

| Task | Rationale |
| --- | --- |
| **(1) Solvation and visualization**: Solvate, visualize, and explain Cytc, selecting a unique molecular representation and explaining their choice. | To assess the participants' foundational skills in using VMD for solvation and visualization tasks, as well as their ability to collaborate with the AI tutor teammate. *Interacting with the AI tutor teammate is assumed to encourage discovery curiosity by prompting participants to select and explain unique molecular representations, sparking curiosity through guided exploration.* |
| **(2) Water molecule calculation:** Calculate the number of water molecules within a specific cutoff distance surrounding different residue types (polar, nonpolar, and charged) of Cytc. | To introduce participants to basic coding tasks that reveal key properties of Cytc, encouraging increased interaction and collaborative problem-solving with the AI tutor teammate. *It is assumed that by adjusting to participants' progress, the AI synchronizes with their pace and cultivates discovery curiosity through iterative guidance, encouraging participants to explore molecular properties within set boundaries.* |
| **(3) Surface area analysis:** Calculate SASA of Cytc and comment on any correlations to the previous water molecule analysis from Task 2. | To encourage participants to develop more complex scripts, AI support and to assess their comprehension by prompting the AI tutor teammate to ask targeted questions that connect Cytc surface area analysis with the insights of subtasks (2). *This may foster team coordination dynamics in a synchronized manner, and, in turn, discovery curiosity.* |
| **(4) Dipole moment calculation:** Calculate and plot the dipole moment of Cytc; comment on its | To challenge participants' problem-solving skills through advanced tasks that require precise coding and critical thinking, facilitated by the AI tutor teammate's guidance on specific, complex questions requiring precise coding |
| functional implications on the function of Cytc. | and reasoning. *It is assumed that the AI tutor teammate aids synchronization by aligning its guidance with participants' progress, stimulating discovery curiosity with questions that demand both technical skills and conceptual exploration of Cytc's dipole moment, linking it to its functional implications.* |

**Research Questions and Hypotheses**

There are two main research questions for this study. The first question (RQ1) is *to what extent and in what ways does an AI-tutor teammate, simulated through a WoZ approach, impact the level of curiosity-driven engagement in students during complex task interactions*? This question assesses whether the AI tutor teammate's curiosity-triggering and curiosity-response strategies effectively stimulate and sustain student curiosity throughout the tasks. We hypothesize that the AI tutor teammate's curiosity-triggering and curiosity-response techniques will maintain and enhance curiosity-driven engagement, leading to increased task involvement and extended interaction duration with the AI tutor teammate ($H_1$). The rationale behind the first hypothesis is that the AI tutor teammate will trigger curiosity (pulling of information) by presenting stimulating facts and posing thought-provoking questions, while it will also respond to the student's inquiries and guide their exploration (pushing of information) in the IMD task. This push-pull information mechanism is designed to maintain a high level of engagement and stimulate deeper cognitive processing. Thus, this mechanism through Human-Computer Interaction (HCI) will harness discovery curiosity as well as learning effectiveness. Building on this first question, which addresses the overall state of student engagement, the second research question seeks to operationalize this engagement by examining its specific behavioral and performance outcomes.

The second question (RQ2) is *to what extent does the use of curiosity-triggering and curiosity-response techniques by the AI-tutor teammate influence the frequency and complexity of student-initiated questions, and what is their overall impact on task performance?* This question explores how specific AI-tutor strategies impact the types of questions students ask, measuring both frequency and depth. We hypothesize that curiosity-triggering and curiosity-response techniques will increase both the frequency and complexity of student-initiated questions, reflecting higher levels of curiosity and cognitive engagement ($H_2$). The rationale for $H_2$ is grounded in research that highlights the relationship between curiosity and the frequency and complexity of questions as indicators of deeper cognitive processing (Abdelghani et al., 2024). When an AI tutor teammate employs curiosity-triggering techniques, e.g., introducing surprising or thought-provoking information, it can spark an innate desire in students to resolve uncertainty and explore further in the IMD task. Similarly, the AI tutor teammate's curiosity-response behaviors, providing elaborative and engaging feedback to students' questions, create an interactive loop that reinforces the value of question-asking as part of the learning process. The AI tutor teammate models a responsive and dynamic learning interaction by validating and expanding students' inquiries, encouraging students to think critically and formulate increasingly complex questions (Abdelghani et al., 2022). This dialogic interaction can deepen students' understanding, often resulting in higher-order questions that reflect curiosity and more sophisticated cognitive processing.

By addressing these two research questions, this study aims to make three key contributions: (1) offering a novel proof-of-concept for an "AI Tutor Teammate" that moves beyond simple question-answering to dynamically regulate curiosity in a complex applied STEM learning environment; (2) introducing and empirically examining a "push-pull" communication dynamic, consisting of curiosity triggers and responses, as a practical framework for operationalizing curiosity theories in AI design; and (3) providing new methodological insights into the relationship between human-AI synchronization and student curiosity by applying nonlinear dynamical systems methods to dyadic coordination.

## METHODS

**Participants**

Recruitment took place via email sent to families during BioSense.Network summer session for K12 students, following Institutional Review Board (IRB) approval. The Center for Research in Education, Science, and Technology (CREST), serving as the study's evaluator, facilitated recruitment and data collection in alignment with grant requirements. Identified participants received an email with a link to the consent form. A total of 11 student participants were primarily recruited from Arizona high schools (between 10th and 12th grades) who attended the Biosense.Network Summer Workshop sessions (between the ages of 13 and 22: $Mean_{Age}$ = 16.57, $SD_{Age}$ = 2.64; Gender: *Male* = 71% and *Female* = 29%). Given the exploratory nature of the study and resource constraints, a smaller sample size was deemed acceptable. However, we recognize this as a limitation affecting statistical power that requires a minimum of 23 participants to achieve .80 power (1 - *β*). Eligible participants were required to speak English, have normal or corrected vision and normal hearing, and be comfortable using a computer mouse and keyboard. They were expected to possess basic computer literacy, sufficient to interact with simulation environments in VMD. Additionally, during summer sessions, these students were introduced to foundational knowledge necessary for the tasks, where they gained a basic understanding of biology (specifically in the area of proteins) and developed initial familiarity with coding in VMD software after approximately two weeks of training in BioSense.Network summer session.

Before the IMD task, a demographics questionnaire was administered. The following key characteristics of the sample included: *Subject Preference:* A majority favored Science subjects (58.3%), followed by Math and Computer Science (33.3% each), *Coding Experience:* A slight majority of participants (57.1%) had prior coding or programming experience from their classes, *Preferred Learning Method:* The most popular choice was "Doing" (hands-on activities) at 71.4%, followed by "Seeing" (e.g., diagrams) and "Reading/Writing" at 57.1% each, *View on Learning Tools:* A large majority (71.4%) believed that using both an AI-Tutor and textbooks together is more effective for learning than using either tool alone.

**Procedure**

Each participant engaged in a 1.5-hour session, detailed in Table 2, consisting of four 15-minute IMD tasks with an AI-tutor teammate on the Zoom platform. Text-based communication was facilitated through Zoom's chat system, while screen sharing allowed participants to view a PowerPoint presentation alongside their progress on the IMD tasks. This setup enabled smooth collaboration with the WoZ "AI-Tutor," providing a comprehensive view of both instructional content and task-related activities. The structured text-chat interactions with the AI tutor teammate, specifically ChatGPT, encouraged participants to explore topics, solve problems, and engage in critical thinking and curiosity-driven exploration.

**Table 2.** The sequence of activities in an experimental session (1.5-hour session).

| Task Order (Duration) |
|---|
| (1) Consent and Briefing (10 min) |
| (2) Demographics (5 min) |
| (3) Curiosity Propensity and Engagement Assessment (10 min) |
| (4) IMD Task 1 (15 min): Solvation and visualization |
| (5) IMD Task 2 (15 min): Water molecule calculation |
| (6) IMD Task 3 (15 min): Surface area analysis |
| (7) IMD Task 4 (15 min): Dipole moment calculation |
| (8) Post-task Curiosity Engagement Assessment (15 min) |
| (9) Debriefing (5 min) |

**Measures**

The measures used to address the research questions are presented in a narrative that progresses to reflect their interconnected nature. First, the measures used to capture the interaction process itself are described; these include the AI's Communication Behaviors (categorized as 'curiosity triggers' and 'curiosity responses') and Dyadic Team Coordination (analyzed via CRQA). Following this, the measures used to assess the impact on student outcomes are detailed. These outcome measures address curiosity-driven engagement (RQ1) through pre- and post-task questionnaires and student questioning and task performance (RQ2) through question complexity coding and a detailed three-part scoring rubric.

*Dyadic team communication.* To assess the varying levels of participant engagement and determine how the AI tutor teammate should adapt to interaction complexity, participants' questions to the AI tutor teammate were categorized into three distinct levels of difficulty (Gobet et al., 2012; King, 1992; Litman & Mussel, 2013). Students' communication behavior with the ChatGPT agent was categorized based on the complexity of their questions, which were classified into three complexity levels, as summarized in Table 3. Additionally, the AI tutor teammate's communication was categorized into two behaviors: curiosity triggers, where the AI posed questions or prompted curiosity, and curiosity responses, where the AI provided detailed explanations or deeper insights.

**Table 3.** Categorizing students' questions with examples by level of complexity.

| Level | Description | Example |
| --- | --- | --- |
| **Basic** | Straightforward requests for information or instructions | "I have downloaded these files and have loaded them into my VMD. What is the next task?" |
| **Intermediate** | Requires more understanding and often involves asking for explanations | "Explain to me how to add a solvation box." |
| **Advanced** | Involves complex, multi-step inquiries or abstract concepts | "Write me a list of '*tk*' console commands to solvate this protein automatically." |

*Dyadic team coordination (communication flow).* To examine how the timely coordination of communication allows an AI to function as a true problem-solving teammate rather than just a reactive tutor, this study analyzed team communication flow as a time series. This time series data set contains multivariate binary measures recorded once at any given time for team members to indicate whether each member sent at least one message (coded as 1) or not (coded as 0). Later, we applied cross-recurrence quantification analysis (CRQA) to the time series data, which was explained in depth later.

*Curiosity propensity and engagement assessment:* Before starting the task, a structured questionnaire adapted from (Litman & Jimerson, 2004) and Tossell et al. (2024) tailored for 10th-grade students was administered to gather background information and assess how often students feel motivated, curious, and engaged with coding challenges, scientific discoveries, and new technological tools. Students responded to a series of items on a 7-point Likert scale, ranging from 1 (almost never) to 7 (almost always), indicating how frequently they felt motivated, curious, or engaged in various scenarios. These items measure two primary dimensions of epistemic curiosity: Deprivation-type curiosity (4 items), which reflects the drive to solve challenging problems and eliminate knowledge gaps, and Discovery-type curiosity (6 items), which reflects the enjoyment of learning new things and exploring unfamiliar subjects. The internal consistency reliability of the scales for the current sample was evaluated using Cronbach's alpha. The overall 10-item scale demonstrated excellent reliability ($\alpha = .93$), while the Deprivation-type subscale ($\alpha = .81$) and the Discovery-type subscale ($\alpha = .87$) both demonstrated good reliability.

*Pre-Post-Task curiosity engagement assessment:* To assess the AI tutor teammate's effectiveness in fostering curiosity, a structured questionnaire that was adapted from Litman and Jimerson (2004) and Tossell et al. (2024) and modified for 10th grade students was administered after the task. This questionnaire specifically evaluated and provided insights into how the AI tutor teammate impacted students' curiosity and engagement with scientific exploration in the IMD tasks conducted through the VMD software platform. Students rated several aspects of their curiosity on a 7-point scale, ranging from 1 (almost never) to 7 (almost always), including their motivation to solve complex VMD problems, desire to explore new VMD tools, and interest in understanding VMD models and results. The instrument's items measure two primary dimensions of epistemic curiosity: Deprivation-type curiosity (4 items) and Discovery-type curiosity (6 items). The internal consistency reliability for this scale was strong, with Cronbach's alpha values indicating excellent reliability for the overall 10-item scale ($\alpha = .91$) and good reliability for the Deprivation-type ($\alpha = .85$) and Discovery-type ($\alpha = .87$) subscales.

*Task performance score.* Task performance was independently evaluated by two experimenters using a detailed scoring rubric encompassing three key criteria, each assessed on a 10-point scale during the experiment and through screen recordings: (1) Task Completion: Assessed the accuracy of completing assigned Interactive Molecular Dynamics (IMD) tasks, with scores ranging from 0 (failure to complete tasks) to 10 (completion of all tasks with high accuracy); (2) Understanding of Concepts: Evaluated the depth of the student's comprehension of the subject matter, with scores from 0 (failure to demonstrate understanding) to 10 (strong understanding with accurate and detailed responses); (3) Engagement and Interaction: Measured active participation and confidence during learning activities, with scores from 0 (no engagement or interaction) to 10 (active engagement with insightful questions and demonstrated confidence). This structured approach comprehensively evaluated each dyad's performance across multiple dimensions.

To assess the consistency of scores between the two independent raters, inter-rater reliability was calculated using a two-way, absolute agreement, single-rater Intraclass Correlation Coefficient (ICC). The analysis indicated excellent reliability for the "Task Completion" score, ICC = .86, 95% CI [.56, .96], $F(9, 9.78) = 13.00$, $p < .001$. The reliability for "Understanding of Concepts" was also good and statistically significant, ICC = .77, 95% CI [.09, .95], $F(9, 3.66) = 13.90$, $p = .015$. The "Engagement and Interaction" score demonstrated moderate and statistically significant reliability, ICC = .56, 95% CI [.00, .87], $F(9, 7.31)$ = 4.62, $p$ = .026.

To explore the relationships between interaction dynamics and learning outcomes, our analytical approach included a comparison of high- and low-performing teams based on total performance scores. Participants were categorized into these groups by splitting the mean task performance score. This division enabled an examination of whether specific communication behaviors, such as asking complex questions, and coordination dynamics, as measured by CRQA metrics, were associated with higher or lower task performance.

## DATA ANALYTICS AND RESULTS

The following results are presented in a progressive narrative because our measures are interconnected and often inform both research questions. First, the results of the overall task performance of the participants, along with their subsequent categorization into high- and low-performing groups, are presented. Then, the findings related to dyadic team coordination and correlation analysis are emphasized to provide key evidence for RQ1 concerning the nature of curiosity-driven engagement. Finally, the student-AI dyadic communication subsection presents the results for our RQ2, which examines the AI's influence on the frequency and complexity of student questioning. This also serves as further behavioral evidence for the engagement discussed under RQ1.

### Task Performance

Participants' performance across task completion, conceptual understanding, and engagement varied, with teams categorized into high-performing ($n$ = 7) and low-performing ($n$ = 6) groups based on the split of the mean total score ($M$ = 24.9; high-performing ≥ 24.9 and low-performing < 24.9). High-performing dyadic teams demonstrated greater alignment between interaction and task completion, as reflected in their higher scores across all dimensions ($M$ = 27.8, $SD$ = 1.73) compared to low-performing dyads ($M$ = 20.5, $SD$ = 3.12). An independent-sample t-test was also conducted to compare the total performance scores for the high- and low-performing groups. There was homogeneity of variances, as assessed by Levene's test for equality of variances, $F(1, 10)$ = 2.66, $p$ = .134. The analysis showed a statistically significant difference in scores between the high-performing and the low-performing teams, $t(10) = 4.90$, $p = .001$, Cohen's $d = 2.97$.

High-performing teams excelled in task completion ($M = 9.60$, $SD = 0.70$), understanding of concepts ($M = 9.40$, $SD = 1.14$), and engagement ($M = 9.80$, $SD = 0.35$). Feedback highlighted active participation and meaningful engagement with the AI tutor teammate, such as asking detailed questions and clarifying task details. For instance, one student participant inquired about the purpose of each line of code in the script, demonstrating curiosity and deeper comprehension, while another consistently asked for alternative strategies to improve their workflow during tasks.

In contrast, low-performing teams scored lower in task completion ($M = 6.20$, $SD = 1.99$), understanding of concepts ($M = 7$, $SD = 0.89$), and engagement ($M = 7.30$, $SD = 1.10$). Feedback suggested that the students in these dyadic teams struggled with hallucinations, such as unclear or misleading AI-tutor's instructions. For instance, one student encountered confusion during protein solvation due to unclear guidance, while another student participant faced misdirection regarding the location of water in a water-free structure. Additionally, the AI's scripting support was often insufficient, preventing participants from completing tasks independently. Some teams engaged with the AI but failed to make significant progress, particularly in complex tasks (e.g., solvation). These findings emphasize the need for more explicit AI guidelines and support to improve task alignment and engagement, especially for low-performing teams. Providing better guidance and task assistance could help minimize performance gaps and enhance learning outcomes.

**Cross-Recurrence Quantification Analysis (CRQA)**

We applied CRQA to each pair's dyadic communication flow data to investigate team coordination dynamics by addressing how each student's coordination with AI is related to one another during a 60-minute interactive molecular task.

*Cross Recurrence Plot (CRP).* The basis of CRQA is the Recurrence Plots (RP; Webber & Zbilut, 1994), which is an illustrative tool for visualizing the temporal evolution of a dynamical system when a system revisits similar states by identifying all pairs of time points in which the system returns to the same state (Coco & Dale, 2014) The cross-recurrence plot (CRP) displays the times when coupled dynamical systems visit similar states. In the current study, we utilize CRQA for categorical (symbolic) time series, with symbols 0 = not messaging and 1 = messaging for each team member. Due to its relative simplicity, we illustrate the concept of recurrence analysis using a CRP; see Figure 2a.

To illustrate recurrence analysis, Figure 2a presents a hypothetical example of a grid of two binary time series of length $N = 10$. In our study, we constructed a time series from communication flow, where '1' indicates a message sent and '0' indicates silence. For instance, if a student is silent in the first time slice (0) and the AI sends a message (1), followed by the student sending a message (1) and the AI going silent (0), the time series begins: the student's sequence is represented along the rows, $n(i) = [0, 1, 0, 1, 1, 0, 1, 0, 1, 1]$, and the AI tutor teammate's sequence is represented along the columns, $p(j) = [1, 0, 0, 1, 0, 1, 0, 1, 1, 0]$. Each $n_{x,i}$ corresponds to the $i^{th}$ message sent by the student ($n$ for student), *and* $p_{y,j}$ represents the $j^{th}$ message sent by the AI tutor teammate ($p$ for AI). $N_x$ and $N_y$ are the total elements that compose the time series for the student and AI tutor teammate, respectively. *If element i and i+1 of the student match with j and j + 1 of the AI tutor teammate, we have matching symbolic elements between the two-time series: ($n_{x,i}$, $n_{x,i+1}$) = ($p_{y,j}$ , $p_{y, j+1}$).* This match can simply be represented by the time index pair ($i, j$), with the understanding that it represents a symbolic "recurrence."

Discrete Cross Recurrence Plots (CRPs) are constructed by placing the student's sequence on the vertical axis and the AI tutor teammate's sequence on the horizontal axis. A recurrence point (value = 1) is marked in a cell whenever a match occurs; otherwise, a non-recurrence point (value = 0) is marked. For instance, the red circulled value at $x(2) = 1$ is repeated at other red circulled $y(2), y(4), y(5), y(7), y(9)$, and $y(10)$, indicating multiple recurrences in the AI's sequence; likewise, the blue circulled value at $x(3) = 0$ is matched at other blue circulled, $y(2), y(3), y(5), y(7)$, and $y(10)$.

For instance, as a fuller illustration, a CRP from one of the pairs (student and AI tutor teammate) is shown in Figure 2b, which indicates points at which matching symbols occur in the two-time series of the student and AI tutor teammate. However, this plot can be significantly larger when constructed from full communication flows in real-world scenarios. The CRP (Figure 2b) visually summarizes the matches

(Figure 2a), illustrating repetitions of 1's and 0's when we build the structure from full message flow; the matches from the left matrix are visually represented as black cells, but their placement appears shifted to the right due to the plot orientation: the x-axis (columns) represents the student sequence, and the y-axis (rows) represents the AI sequence. This shift reflects the reciprocal relationship between the rows and columns, visually mirroring the left matrix. The red and blue circles emphasize that symmetry is maintained because each match in the matrix corresponds directly to a black cell in the plot, preserving the alignment between the two representations.

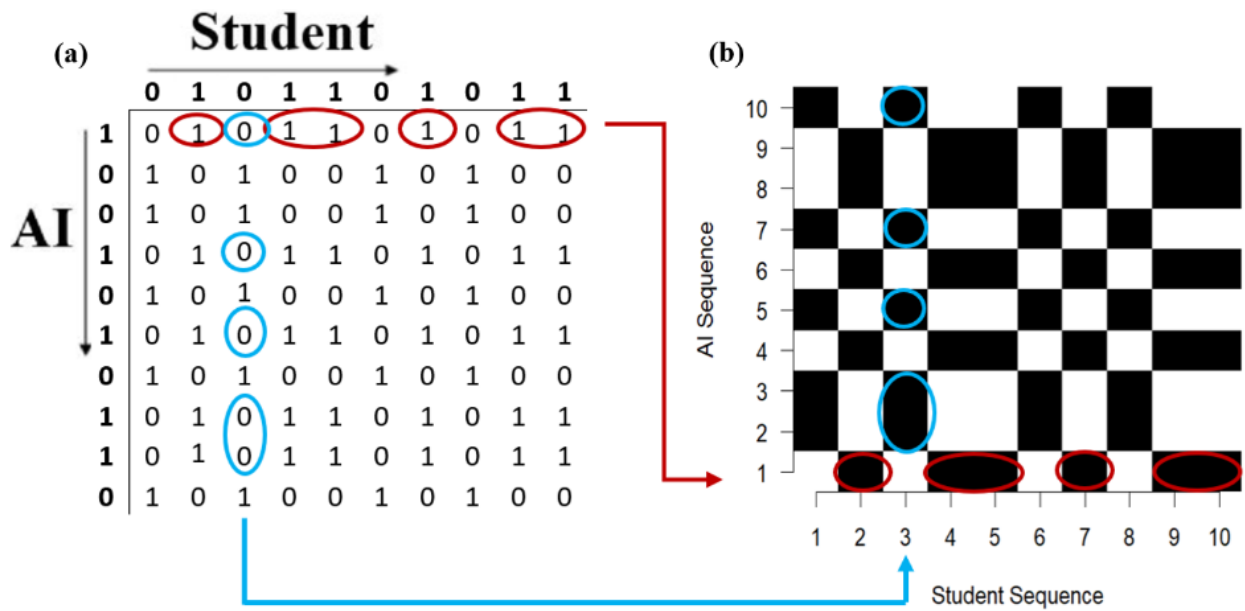

**Figure 2.** Example CRP **(a)** constructed from two-time series. 1's are recorded in cells corresponding to matching elements, with '0' otherwise; **(b)** a full recurrence plot from one of the dyads (student (along the rows) and AI tutor teammate (along the columns)). Filled pixels as symmetry in the plot represent points at which matching elements occur in the two-time series of AI and student.

Figure 3a shows that when a recurrence point is above the diagonal, the time index $j$ is larger than index $i$, or the reverse is true when the point falls below the diagonal line (Dale & Spivey, 2006). We can consider the points above and below the main diagonal line in order to calculate the points. To explore the coordination patterns in these pairings, we conducted a version of CRQA, i.e., compared all time points of two-time series and generated a lag-based percentage of how much matching or "cross-recurring" occurred at each lag. To do so, we created a diagonal-wise recurrence lag profile by plotting this percentage match, known as percentage recurrence (%REC). This profile reflects the coordination pattern between the two-time series (akin to a "categorical" cross-correlation function; for an elegant explanation, see Dale et al. (2011) and Jermann & Nüssli (2011). When the %REC is largely distributed to the right or left of such a plot, it directly affects the leading/ following patterns of the systems producing those time series. Figure 3b demonstrates the recurrence lag profile of the pair of these time series, depicted in Figure 3a as CRP. The profile is constructed by finding how much each time series matches (expressed as percentage recurrence, %REC) when they are lagged relative to one another. To quantify how these profiles changed their position and shape across rounds, we treated the recurrence profiles as distributions of temporal data. The mean lag will be the central tendency of the overall coordination pattern; kurtosis will reflect coordination, and so on (Dale, Warlaumont, et al., 2011). This helps us to describe quantitatively the changes in shape and position that can be seen (Dale, Kirkham, et al., 2011); the example of Figure 3b shows that recurrences were distributed proportionally, i.e., two systems are coupled/ synchronized.

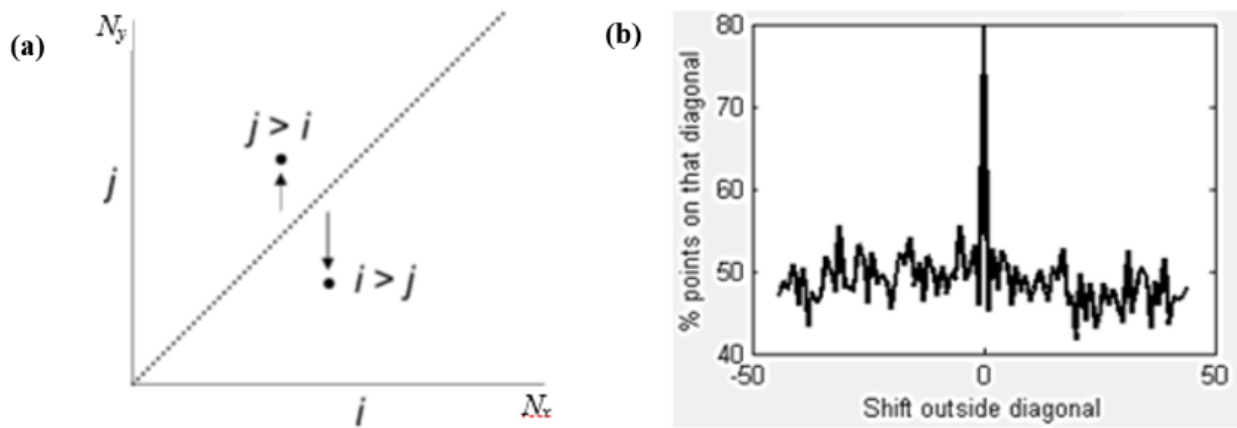

**Figure 3. (a)** A simple diagram of the main diagonal constructed from two-time series and **(b)** the recurrence lag profiles of pairs of two-time series.

*CRQA metrics.* In this study, we utilized the CRP Toolbox (Version 5.29, Release 38; Marwan, 2024) to generate CRPs and their associated metrics, as detailed in Table 4, to analyze the dyadic coordination dynamics between the student and AI-tutor teammate.

**Table 4.** CRQA metrics and their implications for Student-AI-tutor team coordination dynamics.

| Metric | Description | Implication |
|---|---|---|
| **Recurrence Rate (%RR)** | The overall tendency for coupled systems to revisit the same state; Indicates the proportion of recurrent points, showing the amount of information exchange.an index of coupling strength (Bandt et al., 2008) | Represents coupling strength; higher values indicate stronger dyadic coordination in the communication flow. |
| **Determinism (%DET)** | The predictability of coupled systems by evaluating repeating patterns; Reflects the degree of organization; 100% DET indicates perfectly repeating patterns. | Indicates predictability of dyadic coordination; higher values mean more predictable coordination. |
| **Maximum Line Length (MaxL)** | The stability by the longest sequence of recurrent points indicates communication stability, and longer lines suggest more stable interactions, which is an index of attractor stability (Abarbanel, 1996; Stergiou, 2016) | Reflects stability in coordination; higher values indicate stable coordination between the student and AI Tutor Teammate. |
| **Diagonal Recurrence Rate (D%REC)** | The recurrence along diagonals in the plot; Shows synchronous communication behavior. | Represents synchronous coordination, highlighting moments of aligned interaction. |
| **Mean Line Length (MeanL)** | The average length of diagonal lines, excluding isolated points; Reflects the average duration of stable interaction patterns. | Higher values indicate sustained coordination patterns between the student and AI Tutor Teammate. |
| **Trend (UP)** | The trend in recurrence patterns over time (positive direction) shows increasing or stable trends in recurrent patterns over time. | A positive trend indicates an increasing or stable coordination pattern between students and AI Tutor Teammate. |
| **Trend (LO)** | The trend in recurrence patterns over time (negative direction). Shows decreasing trends in recurrent patterns over time. | A negative trend indicates diminishing student and AI coordination and alignment over time. |
| **Entropy** | The complexity or variability in recurrence patterns: Higher entropy reflects more variability and less regularity in the communication flow. | Indicates the unpredictability or diversity in coordination, with higher values representing less routine. |

*Note.* The table is adapted from the following sources (Coco & Dale, 2014; Dale, Kirkham, et al., 2011; Marwan et al., 2007; Webber & Zbilut, 1994).

*Dyadic Team Coordination.* Figure 4 illustrates an example of effective student-AI dyadic team coordination. On Figure 4a, the top red line represents AI speaking activity, and the bottom blue line indicates student speaking activity, where '1' indicates speaking and '0' indicates silence. The CRP (center) shows matching states (black cells) between the student and AI time series, reflecting periods of alignment and interaction patterns. In Figure 4b, the lag profile highlights high synchronization, with a peak at lag 0, indicating strong coordination between the student and AI.

Additionally, CRQA metrics of Figure 4 CRP reveal a highly structured and synchronized coordination between the AI and student, as evidenced by a high %REC (67.30%) and %DET (86.96%). A high D%REC of 58.88% reflects consistent patterns of sequential behavior and synchronized interactions over time. The MaxL = 26 and moderate MeanL = 4.51 mean reflecting stable behavior. The entropy (2.87) suggests a balance between repetitive patterns and variability in dyadic team coordination dynamics. Trends in lagged recurrence reveal a steeper decline when the AI leads (Trend LO = -111.37) than when the student leads (Trend UP = -96.41), indicating a stronger temporal alignment when the student leads the interaction.

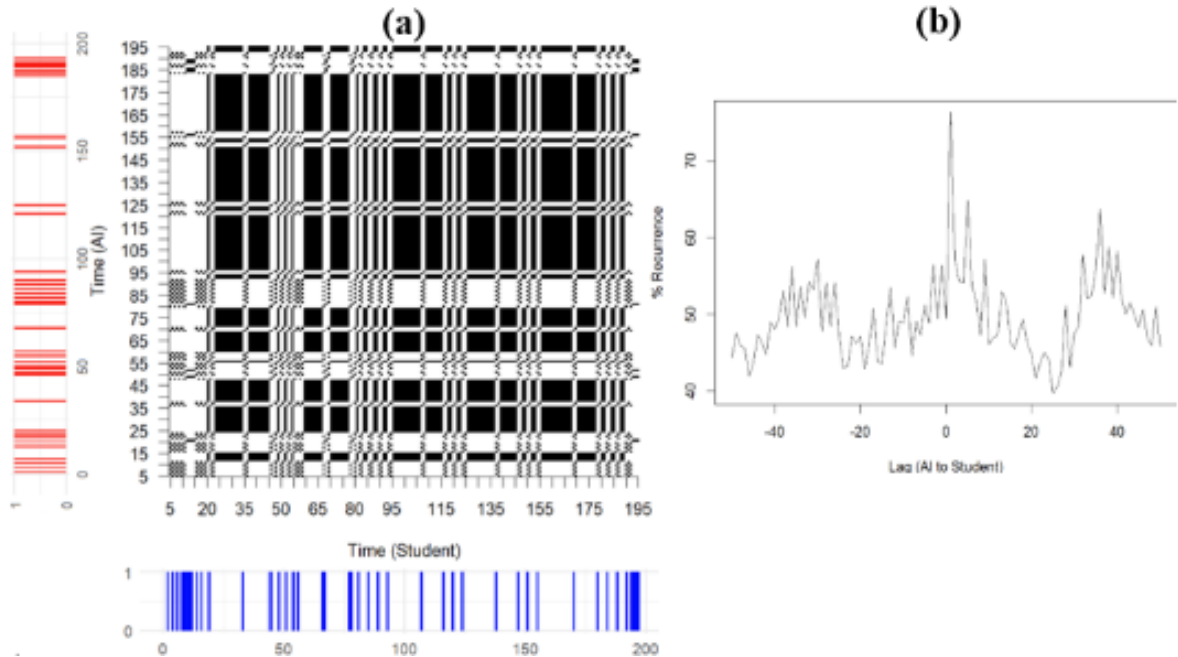

**Figure 4.** High-performing student-AI dyad coordination dynamics: **(a)** Red and blue lines represent AI and student speaking (1 = speaking, 0 = silence). The CRP (center left) shows matching states (black cells), indicating alignment. **(b)** The lag profile means high synchronization, with a peak at lag 0.

Figure 5 illustrates an example of a low-performing student-AI dyadic coordination in which the student was confused during the IMD task, particularly when solvating the protein. The CRP (Figure 5a) shows sparse alignment between the student and AI, reflecting gaps in their interaction. The lag-to-recurrence graph (Figure 5b) further illustrates that engagement peaks when responses are timely, while delays lead to reduced synchronization; i.e., it reveals fluctuating synchronization with low temporal alignment, as indicated by a D%REC of 13.82%. A %REC of 30.07% further supports limited alignment in their activities, likely due to repeated clarification requests from the student. While the high %DET =

78.04% suggests the interaction followed predictable patterns, the negative trends in lag profiles (Trend UP = -78.86, Trend LO = -50.79) indicate diminishing coordination over time.

To examine how the RQA metrics differ between low and high-performing teams, independent sample t-tests were applied. Findings from these tests revealed no statistically significant difference between high and low-performing teams, $p > .05$. The reason for the insignificant results in dyadic team coordination dynamics is that CRQA metrics provide only a high-level overview of team interaction. These metrics may not adequately capture the cognitive and exploratory behaviors, such as asking complex questions or generating curiosity-triggering responses, which are more closely linked to performance outcomes. Therefore, in the next section, we examine the communication behaviors within dyadic teams.

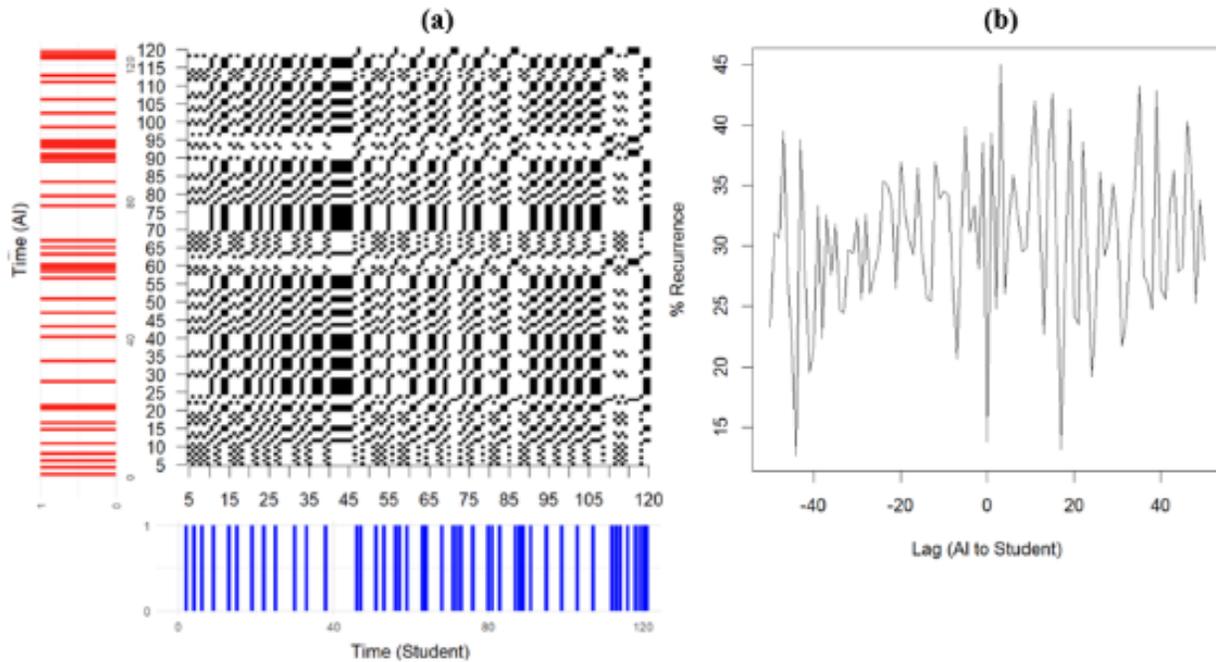

**Figure 5.** Low-performing student-AI dyad coordination dynamics: **(a)** The red and blue lines represent the AI and student speaking frequency, respectively. The CRP (Center) shows matching states (black cells), indicating alignment; **(b)** The lag profile means high synchronization, with a peak at lag 0.

**Dyadic Team Communication**

Students' communication patterns, as measured by question types (basic, moderate, complex) and AI curiosity interactions (responses and triggers), reflect varying levels of engagement with the AI tutor teammate. Based on the descriptive statistics, students asked the level of questions: basic ($M = 4.10$, $SD = $

3.18), intermediate ($M = 3.90$, $SD = 2.28$), and complex ($M = 2.00$, $SD = 2.21$). The cumulative curiosity score aggregates these question types ($M = 10.00$, $SD = 4.55$), indicating moderate interaction complexity. Additionally, the AI tutor teammate exhibited curiosity-triggering behaviors ($M = 8.80$, $SD = 5.83$) and curiosity-response behaviors ($M = 6.10$, $SD = 2.92$), showcasing its dual role in fostering and addressing curiosity during the session.

Figure 6 illustrates smoothed trends of questions by complexity and AI-driven curiosity interactions in the top panels, with cumulative contributions and per-minute occurrences shown in the bottom panels. In Figure 6a (top), the analyzed smoothed trends of questions categorized by complexity show that basic questions (blue) are consistently asked at the beginning but gradually decrease as students gain confidence and move away from simple queries. Intermediate questions (orange) steadily increase and reach a peak midway through the session, reflecting active engagement with moderately challenging aspects of the task as students become more familiar with the simulated IMD task. On the other hand, advanced questions (green) appear later and steadily increase, indicating deeper inquiries as students engage with increasingly complex IMD tasks. The cumulative contribution of complexity levels (bottom panel in Figure 6a) highlights the evolving dominance of question types over time. Basic questions contribute significantly at the beginning of the IMD task, serving as a foundation for learning, but plateau as the session progresses. Intermediate questions dominate the middle phase of the IMD task, underlining students' focus on moderately complex tasks as they gain understanding. Advanced questions contribute substantially in the later phase of the task, illustrating a progressive shift toward in-depth exploration and problem-solving as students' comfort and competence increase.

The AI's curiosity-triggering behaviors (orange, Figure 6b) remain consistent throughout the session, with periodic spikes that align with moments when students likely needed guidance or encouragement. In contrast, curiosity responses (blue, right panels) demonstrate a steady progression, reflecting the AI tutor teammate's adaptive feedback to students' evolving inquiries. The cumulative and per-minute AI behaviors (bottom panel in Figure 6b) reveal that curiosity-triggering messages dominate interaction dynamics, encouraging students to explore further. In contrast, curiosity-response messages support student learning by providing relevant, task-specific explanations.

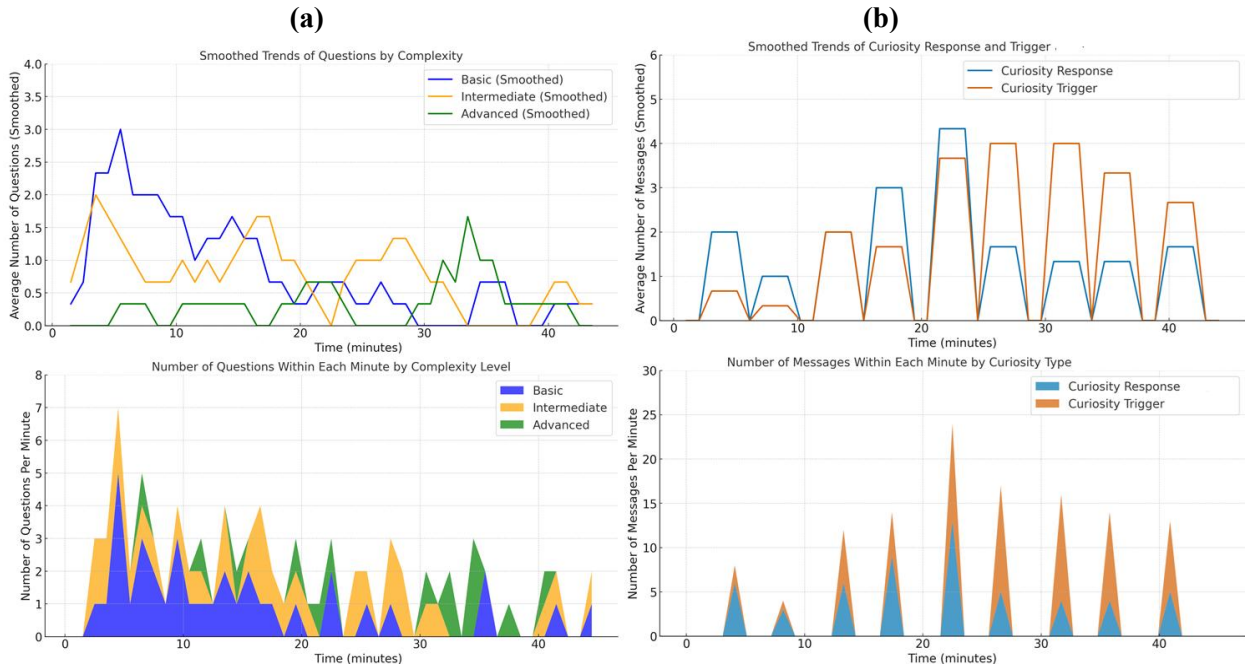

**Figure 6.** Trends and distributions of student question complexity levels (**a**: basic, intermediate, and advanced questions) and AI-tutor's curiosity message types (**b**: curiosity response and curiosity trigger) over a 60-minute session. The top panels show smoothed trends for average activity, while the bottom panels represent the number of occurrences within each minute.

Mann-Whitney U tests, which do not require the assumption of normality, were conducted to examine differences in team communication behaviors between high- and low-performing teams, revealing significant variations in communication patterns. Accordingly, high-performing teams asked significantly more advanced questions than low-performing teams ($U = 21.50, p = .047$) and received significantly more curiosity-triggering messages ($U = 24, p = .014$); see Figure 7. No significant differences were found for basic questions ($U = 13, p = .91$), intermediate questions ($U = 3.50, p = .08$), or curiosity response messages ($U = 10, p = .74$). These results suggest that high-performing teams engage in more complex, curiosity-driven interactions. In contrast, low-performing teams may rely more on intermediate-level queries, indicating different levels of task engagement and depth of understanding.

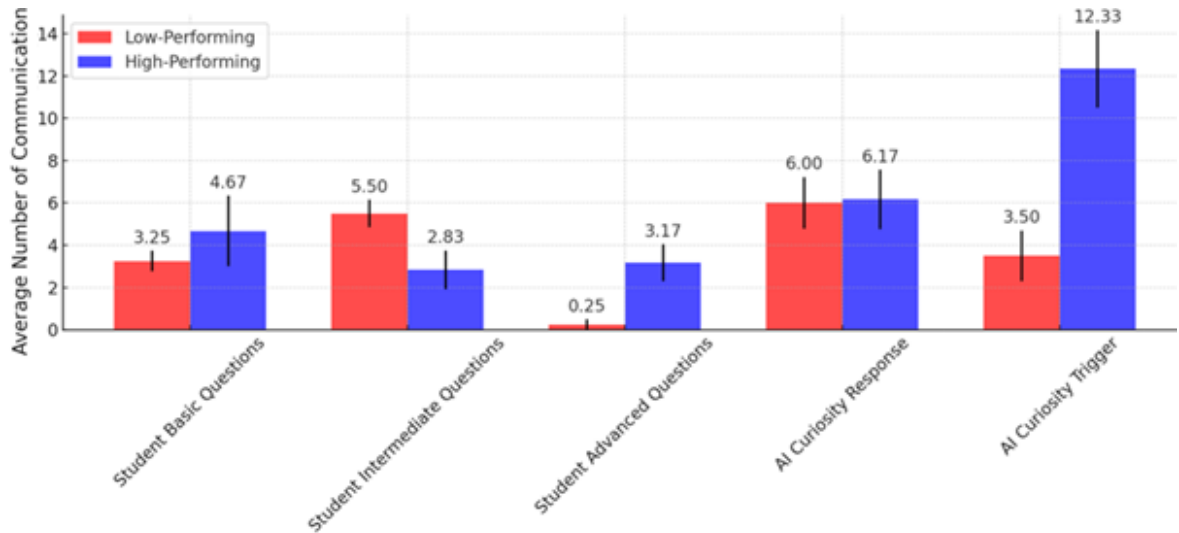

**Figure 7.** The average number of communication behaviors (student questions and AI-tutors response and trigger) across high-and-low-performing dyadic teams.

**Correlation Analysis**

The relationship between curiosity (the sum of three types of student questions) and dyadic coordination metrics was analyzed using Pearson correlation coefficients. For this analysis, the unit of analysis was the individual task completed by each participant, resulting in 44 data points (11 participants × 4 tasks) and yielding 42 degrees of freedom. The results indicate only the following significant relationships: (1) the Diagonal Recurrence Rate (*D%REC*) exhibited a strong positive correlation with curiosity, $r(42) = .69$, $p < .001$. This finding suggests that higher diagonal recurrence, indicative of more structured and predictable interaction sequences, i.e., synchronization, is associated with increased levels of curiosity. Trend.UP showed a moderate negative correlation with curiosity, $r(42) = -.41$, $p < .05$. This indicates that an increase in upward trends, reflecting repetitive patterns or monotony in the interaction, is associated with lower curiosity. The recurrence Rate (*%RR*) was weakly positively correlated with curiosity, $r(42) = .317$, $p = .09$.

**Qualitative Assessment**

To evaluate the strengths and weaknesses of the AI tutor teammate in fostering curiosity, we analyzed mean scores from a curiosity assessment and a pre- and post-task engagement evaluation. The findings were visualized using radar plots to highlight the following areas (Figure 8a): (1) motivation to solve problems, (2) desire to ask more questions, (3) reading about discoveries, (4) eagerness to explore virtual solutions, and (5) seeking more information in interactive molecular modeling. Each area was rated by participants on a scale from 1 to 10. Additionally, specific dimensions of the AI tutor teammate's performance, including accuracy, supportiveness, and reliability, were assessed to identify areas for improvement. These ratings were visually synthesized into a radar plot to compare strengths and weaknesses.

The study evaluated the strengths and weaknesses of an AI tutor teammate in enhancing learning related to VMD through subjective measures derived from participants' responses. The findings highlighted several strengths (Figure 8b), including improvements in understanding VMD models (7/10), fostering interest in biology through molecular visualization (7/10), and stimulating imagination (7/10). Moderate benefits were also observed in participants' ability to consider alternative solutions to VMD problems and improved life sciences knowledge, rated at 6/10. However, the study identified key weaknesses, with the AI tutor teammate's accuracy rating at 4/10, ease of use at 6/10, supportiveness at 5/10, and reliability at 5/10.

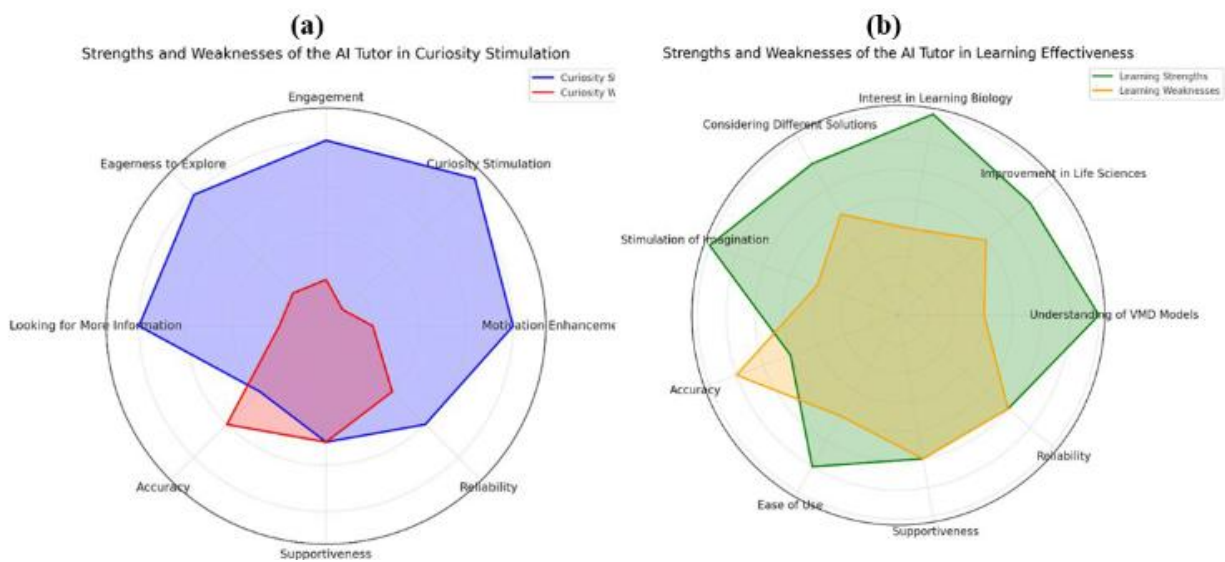

**Figure 8. (a)** Strengths and weaknesses of the AI tutor teammate in curiosity simulation, and **(b)** strengths and weaknesses of the AI tutor teammate in learning effectiveness.

## DISCUSSION

This study outlines a proof-of-concept, AI tutor teammate approach aimed at supporting student curiosity-driven engagement and learning effectiveness; these benchmarks have been shown to be quantifiable using a measurement scheme comprising: a task performance score, a dyadic team coordination metric, a dyadic team communication metric, and pre- and post-curiosity evaluations. Overall findings suggest that while CRQA metrics reflect structural alignment in communication flows (i.e., dyadic team coordination), they are insufficient to explain performance differences in high- and low-performing teams. Instead, these metrics were associated with discovery curiosity, while performance differences are more strongly linked to question complexity and AI-driven curiosity interventions. Capitalizing on the findings related to these two questions, we discuss the requirements for AI design within these interactive sociocognitive contexts (i.e., communication and coordination), particularly in the education of applied mathematics and the life sciences, specifically in teaching applied IMD.

**Enhancing Discovery Curiosity through Team Coordination Dynamics**

In addressing our *first research question on how the AI tutor teammate influenced curiosity-driven engagement*, our findings on team coordination dynamics provide insight into the underlying mechanisms. A WoZ paradigm was utilized to simulate dynamic adjustment of the AI tutor teammate's behaviors, as the current version of ChatGPT-4o does not have the capability to push information, at least for now. This dynamic adjustment helped maintain optimal cognitive disequilibrium, a known driver of curiosity and learning, by modifying the AI tutor teammate's triggers and responses based on the student's performance and engagement levels. The findings indicate that the AI tutor teammate's behaviors, which trigger and respond to discovery curiosity, may stimulate and sustain student interest throughout the IMD tasks. While high-performing teams demonstrated better temporal alignment in their interactions with the AI tutor teammate, low-performing teams experienced weaker synchronization, which contributed to fragmented engagement and less effective task progression.

Although CRQA metrics, such as D%REC (dyadic synchronization) and %RR (coupling strength), did not show significant differences between high- and low-performing teams, these results emphasize D%REC and Trend.UP are crucial predictors of curiosity, indicating that structured and dynamic interactions enhance curiosity, while repetitive experiences may limit it. Overall, the findings highlight a dynamic yet structured, synchronized relationship, where the interaction is predominantly student-driven but retains flexibility and variability.

**Enhancing Discovery Curiosity Through Communication with an AI Tutor Teammate**

Regarding our *second research question, which examined the influence of AI on student questioning*, our findings on communication behaviors suggest that the AI tutor teammate's curiosity-triggering and response behaviors may support curiosity-driven engagement, potentially leading to increased task involvement and longer interaction durations. As illustrated in Figure 6 (left), we observe a pattern where students initially ask basic questions, then appear to progress to intermediate-level inquiries as their understanding deepens, and eventually engage in more advanced-level exploration in IMD tasks. Simultaneously, the AI tutor teammate appeared to exhibit dynamic curiosity-regulation behaviors that may have encouraged exploration and contributed to deeper engagement with the IMD tasks. A previous study by Willett and Demir (2023), which investigated team communication patterns in the Urban Search and Rescue (USAR) task environment, highlights that AI tutor teammates employing metastable behaviors (a balance between intervention frequency and content length) outperformed rigid or highly flexible patterns. This echoes the importance of adaptively calibrating AI interventions to the cognitive and situational demands of other team members, in this case, students.

We observed that high-performing teams appeared to receive more curiosity-triggering interventions from the AI tutor teammate, which may have prompted more frequent engagement with intermediate and advanced-level inquiries compared to low-performing teams. These interventions appeared to support students in transitioning from surface-level questions (i.e., basic questions) to more complex, exploratory ones (i.e., advanced). By potentially introducing cognitive disequilibrium, curiosity triggers may have encouraged students to address knowledge gaps, fostering deeper cognitive engagement and higher-order thinking. This aligns with the push-pull dynamic, where the AI tutor teammate challenges learners with thought-provoking prompts (push) and provides supportive responses to their inquiries (pull), creating a self-reinforcing cycle of discovery. Each successful interaction appeared to enhance metacognitive awareness and intrinsic motivation, potentially driving further exploration. Building on the Information Gap Theory (Berlyne, 1960), the AI tutor teammate's triggers may effectively create knowledge gaps that spark learners' curiosity and motivate them to seek resolution. At the same time, Arousal Theory (Loewenstein, 1994) explains how AI maintains engagement by striking the right balance between challenge and support, stimulating cognitive effort without becoming overwhelming or monotonous. This careful balance appeared to foster flow states during IMD tasks, in which learners seemed fully immersed and motivated. These observations suggest that well-designed, proactive AI interventions may support curiosity and encourage more complex, exploratory questioning (see Figure 6, left).

In contrast, the AI tutor teammate's curiosity-response behaviors were distributed evenly across high- and low-performing teams, primarily as reactive mechanisms (see Figure 7). While these responses clarified questions and resolved confusion, it is possible that they lacked the proactive element necessary to prompt new or more complex questions. However, the interplay between triggers and responses proved essential in influencing the acceleration of student questioning. High-performing teams effectively leveraged this balance, using curiosity triggers to explore and responses to consolidate their understanding. These findings provide initial evidence supporting AI's potential in fostering curiosity-driven interactions, although caution must be exercised due to the limited sample size.

**Enhancing Learning through Team Science**

To address *the final part of our second research question concerning the impact on task performance and learning effectiveness*, as discussed in our theoretical framework, our findings have direct implications for the field of team science, suggesting the AI Tutor Teammate can play a dual role in fostering curiosity and supporting learning. As indicated in the strengths and weaknesses assessment, the AI tutor teammate's intervention strategies appeared to dynamically stimulate curiosity, engaging learners and potentially contributing to their conceptual understanding of IMD tasks. Therefore, the preliminary results from the qualitative assessment suggest the potential of the AI tutor teammate to act as a curiosity stimulant, as indicated by participants' responses shown in Figure 8 (left). The average score given by participants increased in the following categories after their interaction with the AI tutor teammate: Eagerness to explore, engagement with the material, and motivation enhancement. The results demonstrate that the AI tutor teammate effectively bolstered the average participant's eagerness, engagement, and motivation for the subject matter, all of which are markers of curiosity.

Additionally, participants' responses suggest that the AI tutor teammate has the potential to impact the learning effectiveness of participants. For instance, the following categories showed marked improvement: Interest in life science, understanding of VMD models, and consideration for alternate solutions; see Figure 8 (right). The results show that not only did overall comprehension increase, but also the propensity to consider multiple approaches to solve a problem increased, which are factors for effective problem-solving.

Although the qualitative analysis reveals key areas of success for the AI tutor teammate, it also illuminates areas for improvement; notably, participants reported that the AI tutor teammate lacked accuracy, reliability, and supportiveness. For example, one participant noted: *"The code that the agent provided was off base and would throw errors. Prompting it to redo the code yielded the same error"*. Interestingly, this type of failure, while an issue with AI accuracy, can have unintended effects. In this example, the hallucination anecdotally prompted further questioning from the student, which could have incidentally engaged their curiosity as they sought to ensure the AI provided accurate information. In this example, a hallucination, although an issue with AI accuracy, anecdotally prompted further questioning from the student, which could have incidentally engaged their curiosity as they sought to ensure the AI provided accurate information. These results emphasize the student's difficulty following AI tutor teammate instructions, underscoring the need for clearer, step-by-step guidance in similar scenarios. This dynamic interaction between students and AI can foster collaborative learning, allowing for shared knowledge construction and teamwork.

Furthermore, this study highlights the potential of AI tutor teammates as effective teammates in collaborative learning environments. The interplay between students and the AI tutor teammate reflects principles of team cognition, wherein the development of shared mental models through team communication and coordination (Entin & Serfaty, 1999; Stout et al., 1999). The study results suggest a trend where communication complexity increased over time in a synchronized manner, a pattern that appeared to correspond with observations of better team performance, though this relationship requires further investigation. By seamlessly integrating scaffolding with autonomy, the AI tutor demonstrated its capacity to optimize engagement and learning through the lens of communication content and flow (coordination), underscoring its potential as a transformative tool in educational settings. This synergy in learning aligns with research on human-AI teaming (HAT), emphasizing the importance of mutual adjustments, goal alignment, and adaptability in achieving task success (Demir & Cooke, 2014). Furthermore, the AI tutor teammate's ability to scaffold learning while maintaining engagement parallels the functions of human teammates in scientific collaborations, where balancing autonomy and support is critical for fostering curiosity.

Building on this foundation of team cognition, the AI tutor teammate appeared to support learning by operating within students' zone of proximal development (ZPD)—the gap between what a learner can achieve independently and what they can accomplish with targeted interventions (Vygotsky, 1980). By introducing challenging yet achievable tasks and providing scaffolding through timely feedback, the AI tutor teammate may have helped create an environment where learners could actively construct their understanding of complex concepts, as suggested by students asking more complex questions over time. This guided discovery approach may have contributed to deeper conceptual understanding and promoted autonomy, allowing learners to engage critically and build upon prior knowledge. The feedback loop between the AI and the participants appeared to dynamically adjust support to their evolving needs, potentially fostering both confidence and competence in mastering IMD.

Findings on task performance further highlighted the AI tutor teammate's impact on engagement and learning outcomes. Real-time feedback and guidance kept learners actively involved in tasks, leveraging principles of distributed cognition, where knowledge is shared and co-constructed between the AI and the learner. This partnership may result in enhanced problem-solving and task completion by distributing cognitive demands, enabling participants to tackle more complex challenges with sustained focus by effectively coupling and synchronizing. By seamlessly integrating scaffolding with autonomy, the AI tutor teammate demonstrated its capacity to optimize engagement and learning through the lens of communication content and flow (coordination), underscoring its potential as a transformative tool in educational settings.

**Transforming Education through AI-Teammate**

The findings suggest the transformative potential of AI-powered curiosity stimulation in K-12 STEM education, in this case, high-school students, offering preliminary insights into creating personalized learning environments that foster deeper engagement and understanding. Our findings support the idea, grounded in our theoretical framework, that by helping to manage cognitive conflict and resolution, interactive AI systems can effectively support learners as they navigate complex concepts. The AI tutor teammate's observed capacity to dynamically calibrate challenges and provide timely support suggests possibilities for how technology could potentially extend traditional scaffolding approaches, bridging knowledge gaps while maintaining learners' motivation and focus. This approach aligns with modern educational goals of promoting critical thinking, problem-solving, and self-directed learning in STEM disciplines.

For the design of adaptive AI tutor teammates, the importance of balancing push-pull dynamics in curiosity-driven learning environments is paramount. Effective systems must stimulate exploration (push) while providing structured guidance (pull) to help learners progress without diminishing their autonomy. This dynamic balance is intended to help learners remain active participants in their educational journey, constructing knowledge through guided discovery rather than passively receiving information. These findings highlight the need for AI systems to adapt to individual learner profiles, leveraging real-time assessments to create personalized pathways through complex material while fostering curiosity and engagement. Such advancements, if further developed and validated through extensive research, hold the potential to significantly contribute to enhancing K-12 STEM education by helping to make learning more accessible, meaningful, and tailored to the diverse needs of students.

**Limitations**

This study employed a WoZ setup, which, while effective in simulating adaptive AI tutor teammate behaviors, introduces limitations in emulating a fully autonomous AI system. The reliance on experimenter intervention in the WoZ framework may have influenced interaction patterns, potentially introducing biases affecting findings' generalizability to real-world, autonomous AI applications. As an exploratory study, the the research has a small sample size that limits its statistical power, and reduces the ability to generalize results to broader populations. This limitation was particularly evident in the CRQA metrics analysis, where high variability across teams may have obscured significant differences despite observed trends in synchronization dynamics from an exploratory perspective.

In contrast, coding for question types and the AI tutor teammate's questions and responses exhibited less variability and revealed more explicit patterns tied to team performance. The sample size and effect sizes for CRQA metrics may have been insufficient to achieve statistical significance, underscoring the need for larger-scale studies to validate and expand these findings.

**Future Directions**

*Development of adaptive AI-tutor systems.* Future research should prioritize the development of adaptive AI tutor teammate systems capable of flexible interventions (i.e., curiosity-triggering and response), allowing them to tailor interaction strategies in real-time based on assessments of student engagement and cognitive load. Such systems would recognize and respond to discovery curiosity, from initial situational interest to more maintained epistemic engagement. The integration of multimodal sensor-based technologies, such as eye-tracking, facial expression, physiological monitoring, or sentiment analysis, could enable these AI systems to detect shifts in students' curiosity levels and dynamically adjust their behavior to maintain engagement. Further exploration of these metrics will help inform the design of AI tutor teammates that can enhance discovery curiosity and optimize learning outcomes.

*Testing diverse AI conditions.* It is essential to investigate how the predictability of AI tutor teammates' behavior and users' trust in AI impact discovery curiosity and learning effectiveness. This can be achieved by testing various conditions influenced by AI tutor teammate system interventions, such as highly predictable AI tutor teammates or those exhibiting abnormal behaviors (e.g., an AI tutor teammate that demonstrates occasional anomalies in its communication). Additionally, exploring AI tutor teammates that use analogy and metaphorical thinking for better explanations should be considered. It is important to note that experiments should also include no-AI control groups as a baseline. These studies could mirror different pedagogical approaches, ranging from structured guidance to discovery-based learning, offering insights into how AI can adapt to different teaching contexts. Moreover, examining more complex human-AI interaction factors, e.g., the timing of responses and the depth of information provided, could enhance our understanding of how to design AI systems that foster engagement and improve educational outcomes.

*Integration into classroom environments.* The integration of AI tutor teammates into classroom environments should be studied using frameworks such as TPACK (Technological Pedagogical Content Knowledge). This research could assess how AI systems function as tutors and catalysts for innovative teaching practices. AI tutor teammates could be designed to adjust task complexity and support levels dynamically, enabling students to enter and sustain flow states during learning. Pilot studies should explore how these systems can support teachers in managing complex educational tasks and balancing personalized learning for students with the broader instructional needs of the classroom.

*Analysis of multimodal interaction data.* Future studies should delve into human-AI team interaction's verbal, nonverbal, and emotional dimensions to uncover how these elements influence curiosity-driven learning. Advanced analytics, such as machine learning models, multimodal data fusion techniques, and multimodal nonlinear dynamical systems methods, can help analyze patterns in multilevel interaction data and reveal critical factors that drive curiosity and engagement. These insights would enable the development of more refined AI systems that are better equipped to adapt to individual and group dynamics, ultimately improving the effectiveness of AI-driven educational interventions. These directions highlight the need for interdisciplinary approaches, integrating advancements in AI, education, human factors, and cognitive science to realize the full potential of AI tutor teammates in fostering curiosity and enhancing learning experiences.

## CONCLUSION

The primary contribution of this exploratory study is to provide a proof-of-concept framework for how AI tutor teammates can be designed to foster discovery curiosity in complex STEM domains. Our findings provide initial indications that AI tutor teammates may have a role in this process, specifically by leveraging adaptive curiosity-triggering and response behaviors to enhance engagement and support meaningful knowledge construction in IMD tasks. The interplay of these triggers and responses, encapsulated in a push-pull dynamic, appeared to be an important dynamic that could potentially contribute to fostering metacognitive growth and sustaining intrinsic motivation.

This exploratory study involving 11 high school students provides initial indications that AI tutor teammates may have a role in fostering discovery curiosity, and supporting learning within specific STEM contexts, suggesting a promising avenue for more extensive research. By leveraging adaptive curiosity-triggering and response behaviors, the AI tutor teammate appeared to play a notable role in contributing to enhancing engagement, promoting question complexity from basic to advanced, and supporting the construction of meaningful knowledge in IMD tasks. In high-performing teams, there were indications that curiosity-triggering interventions were associated with creating cognitive disequilibrium, which potentially motivated learners to resolve knowledge gaps and drive deeper exploration. The interplay of curiosity triggers and responses, encapsulated in a push-pull dynamic, appeared to be an important dynamic that could potentially contribute to fostering metacognitive growth and sustaining intrinsic motivation.

Synchronization between the AI tutor teammate and students further highlighted the importance of alignment in human-AI interactions. Patterns of communication stability and the development of shared mental models are theoretically linked to task performance, and our findings are consistent with the idea that effective human-AI teaming may depend on the ability to dynamically coordinate behaviors and goals, which supports previous studies (e.g., Demir & Cooke, 2022). While structural synchronization metrics revealed trends in alignment, their lack of statistical significance points to the need for a sufficiently large sample size to achieve sufficient power and better understand how synchronization (or CRQA metrics) impacts learning outcomes.

The findings also suggest the potential value of AI tutor teammates in contributing to the creation of personalized learning environments within learners' zones of proximal development. The AI tutor teammate, as designed in this study, aimed to build upon traditional educational approaches by providing tailored scaffolding and adjusting challenge levels, thereby potentially supporting the progression from surface-level learning to advanced conceptual understanding. This adaptability further suggests the potential for AI systems to contribute to the evolution of STEM education by promoting curiosity-driven exploration and active knowledge construction, although this potential requires significant further research and development.

**Author Contribution**

M.D. wrote the main manuscript text, conducted visualization, and performed methodology, formal analysis, and data curation. J.M. reviewed and edited the manuscript, as well as contributed to data analysis and data curation. P.M., J.N., C.K.C., and A.S. provided additional review and editing. All authors reviewed and approved the final manuscript.

**Declaration of competing interest**

The authors declare that they have no known competing financial interests or personal relationships that could have appeared to influence the work reported in this paper.

**Funding**

This work was partially funded by the National Defense Education Program (NDEP) for Science, Technology, Engineering, and Mathematics (STEM) Education, Outreach, and Workforce Initiative Programs under Grant No. HQ0034-21-S-F001. It was also funded by the Air Force Office of Scientific Research (AFOSR) for STEM Education under Grant No. FA9550-23-1-0632.

**Data Availability**

Data will be made available on request.